\documentclass[12pt,preprint]{aastex}

\def\etal       {et al.}
\def\ie         {{i.e.},\ }
\def\eg         {{e.g.},\ }

\def\sixsig     {$6\sigma$}

\def\sgra    {Sgr~A*}
\def\msun    {M$_\odot$}

\def\Vorb    {\ifmmode {V_{orb}} \else ${V_{orb}}$ \fi}
\def\Vdot    {\ifmmode {\.V} \else ${\.V}$ \fi}

\def\arcm{\ifmmode {' }\else $' $\fi}
\def\arcs{\ifmmode {'' }\else $'' $\fi}
\def\arcmper{\ifmmode \rlap.{'} \else $\rlap{.}' $\fi}
\def\arcsper{\ifmmode \rlap.{''} \else $\rlap{.}'' $\fi}
\def\porm   {\ifmmode\pm\else$\pm$\fi}
\def\kms    {\ifmmode{{\rm ~km~s}^{-1}}\else{~km~s$^{-1}$}\fi}
\def\masy   {\ifmmode{{\rm mas~y}^{-1}}\else{mas~y$^{-1}$}\fi}
\def\micron {\ifmmode{\mu{\rm m}}\else{$\mu$m}\fi}
\def\a      {\ifmmode {\rlap.}^{''}\! \else ${\rlap.}^{''}\!$\fi}
\def\p      {\phantom{0}}

\newbox\grsign \setbox\grsign=\hbox{$>$} \newdimen\grdimen \grdimen=\ht\grsign
\newbox\laxbox \newbox\gaxbox
\setbox\gaxbox=\hbox{\raise.5ex\hbox{$>$}\llap
     {\lower.5ex\hbox{$\sim$}}}\ht1=\grdimen\dp1=0pt
\setbox\laxbox=\hbox{\raise.5ex\hbox{$<$}\llap
     {\lower.5ex\hbox{$\sim$}}}\ht2=\grdimen\dp2=0pt

\def\lax{\mathrel{\copy\laxbox}}

\slugcomment{2006 December 4}

\shorttitle{\sgra\ Position: III}
\shortauthors{Reid \etal      }

\begin{document}

\title{The Position of Sagittarius A*: \\ 
       III. Motion of the Stellar Cusp}

\author{M. J. Reid}
\affil{Harvard--Smithsonian Center for Astrophysics,
    60 Garden Street, Cambridge, MA 02138}
\email{reid@cfa.harvard.edu}

\author{K. M. Menten}
\affil{Max-Planck-Institut f\"ur Radioastronomie,
       Auf dem H\"ugel 69, D-53121 Bonn, Germany}
\email{kmenten@mpifr-bonn.mpg.de}

\author{S. Trippe, T. Ott \& R. Genzel}
\affil{Max-Planck-Institut f\"ur extraterrestrische Physik,
       Postfach 1312, D-85741 Garching, Germany}
\email{trippe@mpe.mpg.de, ott@mpe.mpg.de, genzel@mpe.mpg.de}

\begin{abstract}

In the first two papers of this series, we determined the position of 
\sgra\ on infrared images, by aligning the positions of red giant stars
with positions measured at radio wavelengths for their circumstellar
SiO masers.  In this paper, we report detections of 5 new
stellar SiO masers within 50'' (2 pc) of \sgra\ and new and/or 
improved positions and proper motions of 15 stellar SiO masers.  
The current accuracies are $\approx1$~mas in position and 
$\approx0.3$~\masy\ in proper motion.  We find that the proper motion
of the central stellar cluster with respect to \sgra\ is less than 
$45$~\kms.  
One star, IRS~9, has a three-dimensional speed of $\approx370$~\kms\ 
at a projected distance of 0.33~pc from \sgra.  
If IRS~9 is bound to the inner parsec, this requires an enclosed mass 
that exceeds current estimates of the sum of the mass of \sgra\ and  
luminous stars in the stellar cusp by $\approx0.8\times10^6$~\msun.
Possible explanations include i) that IRS~9 is not bound to the central
parsec and has ``fallen'' from a radius greater than 9~pc, ii) that
a cluster of dark stellar remnants accounts for some of the excess mass,
and/or iii) that R$_0$ is considerably greater than 8~kpc.

\end{abstract}


\section{Introduction}

Sagittarius A* (\sgra), the compact radio source at the center of the Galaxy 
\citep{BB74}, is almost surely a super-massive black hole.  Infrared-bright
stars orbit about the radio position of \sgra\ and require 
$\approx3.9\times10^6$~\msun\ within a radius of $\approx50$~AU 
\citep{E05,G05}.  
While orbiting stars move at thousands of \kms, \sgra\ is essentially 
stationary at the dynamical center of the Galaxy, moving $<2$~\kms\ out of 
the Galactic plane, \citep{BS99,R99,RB04}.  This indicates that the
radiative source \sgra, which is less than 1~AU in size 
\citep{Rogers04,K98,D01,B04,S05}, 
contains a significant fraction ($>10\%$) of the gravitational mass \citep{RB04}.

The Galactic Center stellar cluster contains red giant stars 
that are both strong radio sources (from circumstellar SiO
maser emission) and bright infrared sources.  Because these stars
are visible at both radio and infrared wavelengths, one can use their
radio positions, measured with respect to \sgra, to determine accurate 
scale, rotation, and distortion corrections for an infrared image.
This allows the highly accurate radio reference frame to be transfered to
the infrared images, improving the quality of the infrared astrometry and
precisely locating the position of \sgra.
In \citet{MREG97} and \citet{R03}, hereafter Papers I \& II, 
we developed this technique and determined the position of \sgra\ 
on diffraction-limited 2.2~\micron\ wavelength images of the Galactic
Center with an accuracy of $\approx15$~mas.
Locating \sgra\ on infrared images has been important for
determining its emission in the presence of confusing stellar sources and
verifying that foci of stellar orbits coincide with the position 
of \sgra.  This links the {\it radiative} (compact radio) 
source with the {\it gravitational} source.

We present new VLA observations of stellar SiO masers in the central cluster,
updating their positions and, by more than doubling the observing time-span, 
significantly improving their proper motion determinations.  
We place the proper motions of infrared
stars in the central cluster in a reference frame tied to \sgra.
In \S2 we describe the radio measurements of the positions and
proper motions of 15 SiO maser stars, and in \S3 we present the latest
infrared positions and proper motions of those stars within 20\arcs\ 
of \sgra.   We use these data to transfer the infrared 
proper motions to a reference frame tied to \sgra\ in \S4.
Finally, in \S5, we use the three-dimensional speeds and projected
offsets of stars from \sgra\ to constrain the combined mass of \sgra\ and
the central (luminous and dark) stellar cluster.

\section{Radio Observations}

Over the period 1995 to 2006 we have searched for and mapped
SiO masers associated with late-type stars that are projected near \sgra.  
We used the NRAO\footnote{NRAO is a facility of the National Science Foundation
operated under cooperative agreement by Associated Universities, Inc.} 
VLBA and VLA to measure accurately the relative positions of SiO maser 
stars and \sgra.  

Red giant and super-giant stars of late-M spectral class often exhibit
SiO masers in their extended atmospheres.  These masers
are strongly variable over time scales of $\sim1$~y.
SiO maser emission emanates from a typical radius of 
$\sim4$~AU (\eg \citet{DK03}), which corresponds to $\sim0.5$~mas 
at the assumed 8.0~kpc distance of the Galactic Center \citep{R93}.  
As our measurements were made over a time-span 
much longer than the stellar cycle, we cannot track individual maser features,
and we accept an intrinsic stellar position uncertainty of about $\pm0.5$~mas 
owing to possible variations across the maser shell.  
For a late-type super-giant, the stellar radius is
considerably larger than for a red giant of similar spectral class, and
its SiO masers are found at radii roughly an order of magnitude larger.  
Variation of the SiO masers in a super-giant can considerably degrade 
inferred stellar position and proper motion measurements
(see discussion of IRS~7 in \S2.1).

\subsection{VLA Observations}

Our VLA observations were conducted in the A-configuration in 1995 June 
(reported in Paper I), 1998 May and 2000 October/November (reported in Paper II)
and 2006 March (reported here).  
Near the Galactic Center, SiO masers are likely detectable 
over a very wide range of velocities, probably exceeding 700~\kms.
However, wide-band observations at the VLA are currently severely
limited by the correlator.  In order to obtain adequate spectral
resolution and sensitivity, we chose to limit our velocity coverage and 
observe in seven 6.25~MHz bands 
(each covering $\approx40$~\kms\ excluding band edges).   
We observed in both right and left circular polarization for each band
and obtained 64 spectral channels per band,
resulting in uniformly weighted spectral channel spacings of 
about 98~kHz or 0.67~\kms.  

Our 2006 observations were conducted on March 5, 18, \& 19.
The VLA had 24 antennas in operation, and the synthesized beam toward 
\sgra\ was about $86 \times 33$~mas elongated in the north-south direction.  
Except for occasional calibration sources, we pointed on the position
of \sgra, allowing detection of masers within the primary beam of a
VLA antenna ($\approx30''$ HWHM at 43 GHz).
We observed by cycling among bands centered 
at LSR velocities of $-346, -111, -73, -39, -1, +40~{\rm and}~+75$\kms;
the latter six bands covered the LSR velocity range 
$-131~{\rm to}~+95$\kms\ with only two small gaps.
This setup allowed for deeper integrations for the previously
known, or suspected, SiO maser stars than would be possible for a
``wide-open'' search.

Initial calibration of the VLA data was done in a standard manner 
recommended by the Astronomical Image Processing System (AIPS) documentation.
The flux density scale was based on observations of 3C~286, 
assuming 1.49~Jy for interferometer baselines shorter than
300~k$\lambda$.  Amplitude and bandpass calibration was accomplished
with observations of NRAO~530, which had a flux density of 2.8~Jy
during the 2006 observations.
The visibility data were then self-calibrated (amplitude and phase) on \sgra\
for each individual 10-sec integration.  This removed essentially all
interferometer phase fluctuations, owing to propagation through the 
Earth's atmosphere, and placed the phase center at the position of \sgra.

\sgra\ is known to vary occassionally in flux density by $\approx$ 10\% 
on hourly time scales \citep{YZ06}.  Were this to happen during our
observations, self-calibrating on \sgra\ would introduce false gain 
(amplitude) variations of a similar magnitude and degrade somewhat 
the dynamic range of the SiO maser images.  However, this is
unlikely to shift the measured position of the masers significantly,
as the positions are determined almost entirely by the phase data.

We searched for new maser stars by making very large images, covering 
about $\pm50$\arcs, or most of the primary beam of an individual 
VLA antenna at 43 GHz, about \sgra.  This was done by limiting the range of
uv-data used and resulted in maps with a nearly circular beam of 0.15\arcs.
Typical rms noise levels in these images were near 9~mJy, allowing
\sixsig\ detections of 54~mJy.  Five new SiO maser stars were discovered:
SiO-14, SiO-15, SiO-16, SiO-17, and IRS 19NW.

\begin{figure}

\epsscale{0.95}
\plotone{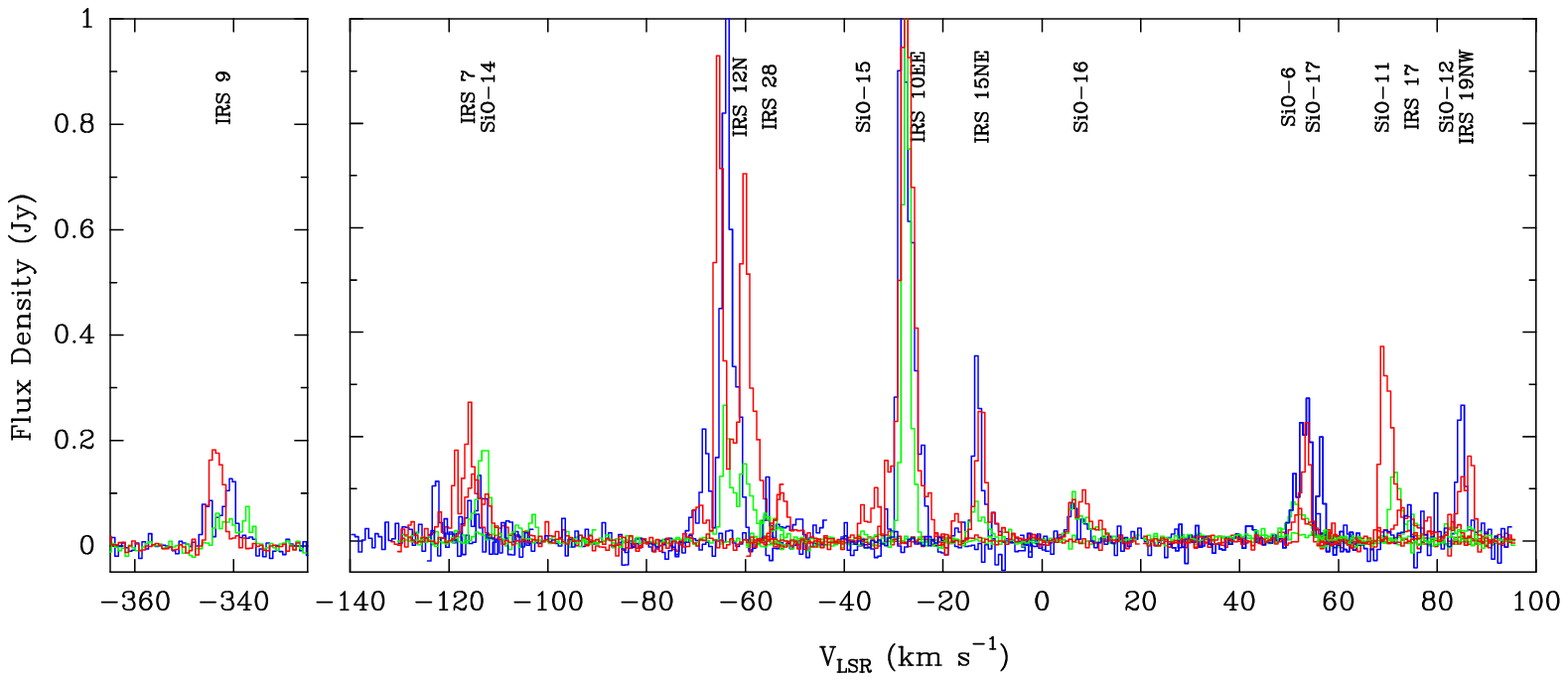}
\caption{Composite spectrum of stellar SiO masers detected with the VLA in 
1998 ({\it blue}), 2000 ({\it green}) and  2006 ({\it red}) 
observations.  Stars are identified at the
top of the spectrum at their approximate stellar velocities.   
            \label{fig:giant_spectrum}
        }
\end{figure}

Once the approximate location of a maser was known, 
either from previous observations or from the large images, 
we mapped each band with up to five small sub-images centered 
on known or suspected masers with emission in that band.  
These synthesized maps typically had single spectral-channel noise levels
of about 5~mJy.   
We always included a sub-image for \sgra\ at the phase 
center of the interferometric data.  By simultaneously imaging the
stellar SiO masers and the continuum emission from \sgra, 
the strong continuum emission from \sgra\ did not degrade the
detections of relatively weak SiO masers far from \sgra.
A composite spectrum from all seven observing bands
for the 1998, 2000, and 2006 observations is shown in 
Fig.~\ref{fig:giant_spectrum}.

As described in Paper~II, we obtained a single position for each star at 
each observing epoch by 1) fitting a 2-dimensional Gaussian brightness 
distribution to each spectral channel with detectable SiO emission,
2) averaging, using variance weighting, to obtain a best stellar position 
and estimated uncertainty (if the reduced $\chi^2$ was greater than unity, 
we increased the position uncertainties accordingly), and 3) correcting for
differential aberration, an effect of $<1$~mas for stars $<15$~\arcs\ 
of \sgra.
   
We list the positions of the SiO maser stars, relative to \sgra, in
Table~\ref{table:positions}.  
We include the results from the 1995 VLA observations reported 
in Paper I, the 1998 and 2000 VLA observations reported in Paper II,
and the 2006 results reported in this paper. 
Since stellar SiO masers are variable in
strength over the period of the stellar pulsation 
and the sensitivity of each epoch's data differed somewhat, only the
stronger sources are detected at all epochs.

One of the stars, IRS~7, is a super-giant.  As discussed in Paper~II,
it has SiO maser features spread over $\approx20$~\kms\ and is
highly variable.  We would expect its stronger SiO maser peaks could be
spread over a region of at least $10$~mas.
Thus, the positions determined from the brightest SiO maser feature(s) 
in IRS~7 may not indicate the stellar position to better than about 5~mas, 
and we have increased the position uncertainties for IRS~7  
in Table~\ref{table:radio_motions} to allow for this possibility.

\begin{figure}
\epsscale{0.85}
\plotone {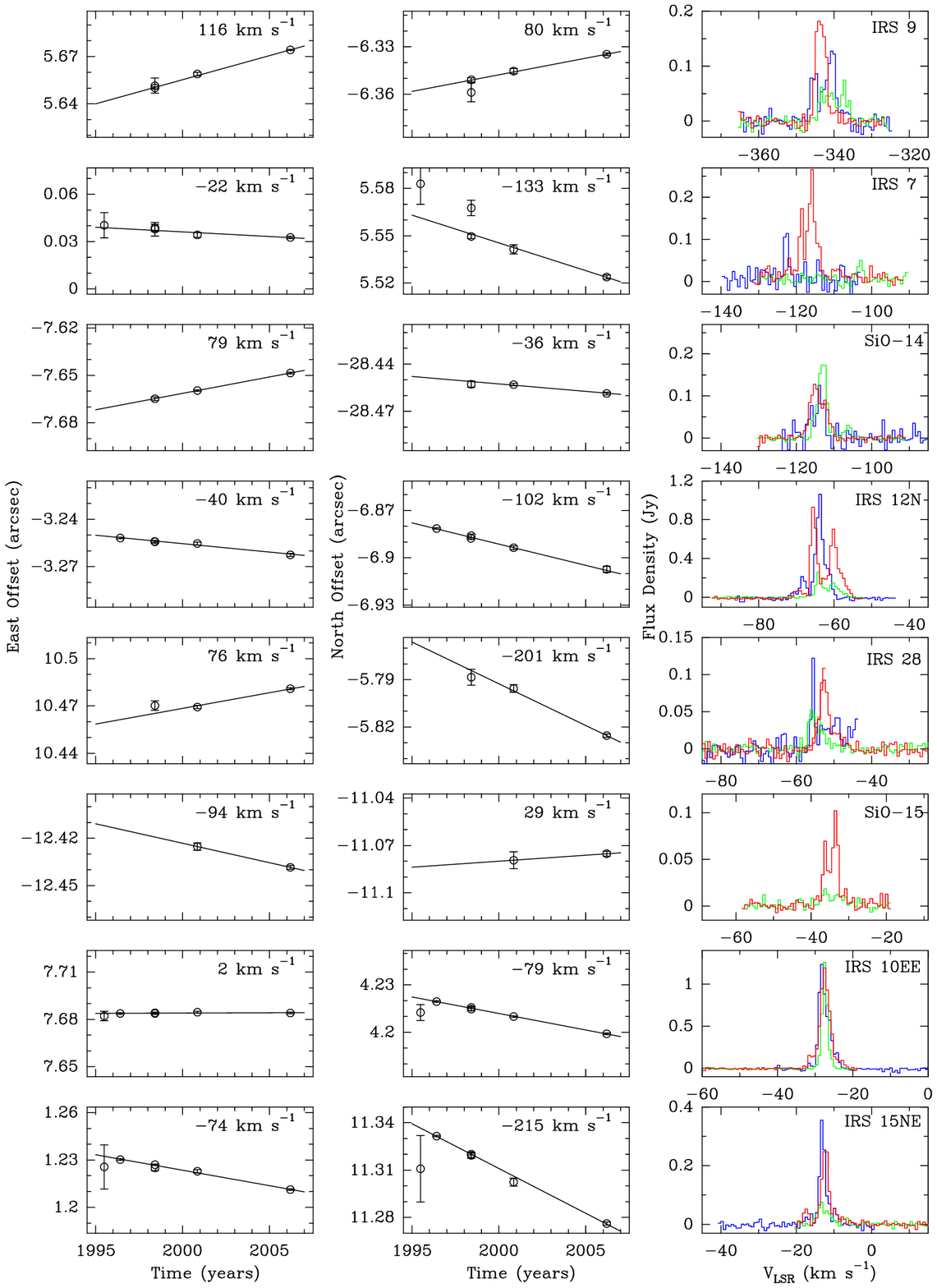} 
\caption{Eastward (left panels) and northward (middle panels) position
offsets from \sgra\ versus time for the eight SiO maser stars with
negative LSR velocities.  Solid lines are variance-weighted best-fit 
proper motions.  
The linear speed is indicated in each frame, assuming a distance of 8.0~kpc.  
Spectra from 1998 ({\it blue}), 2000 ({\it green}) and  
2006 ({\it red}) observations are also shown (right panels). 
Star names are indicated in the upper
right corner of the right panels.
            \label{fig:motions_a}
        }
\end{figure}

\begin{figure}
\epsscale{0.85}
\plotone {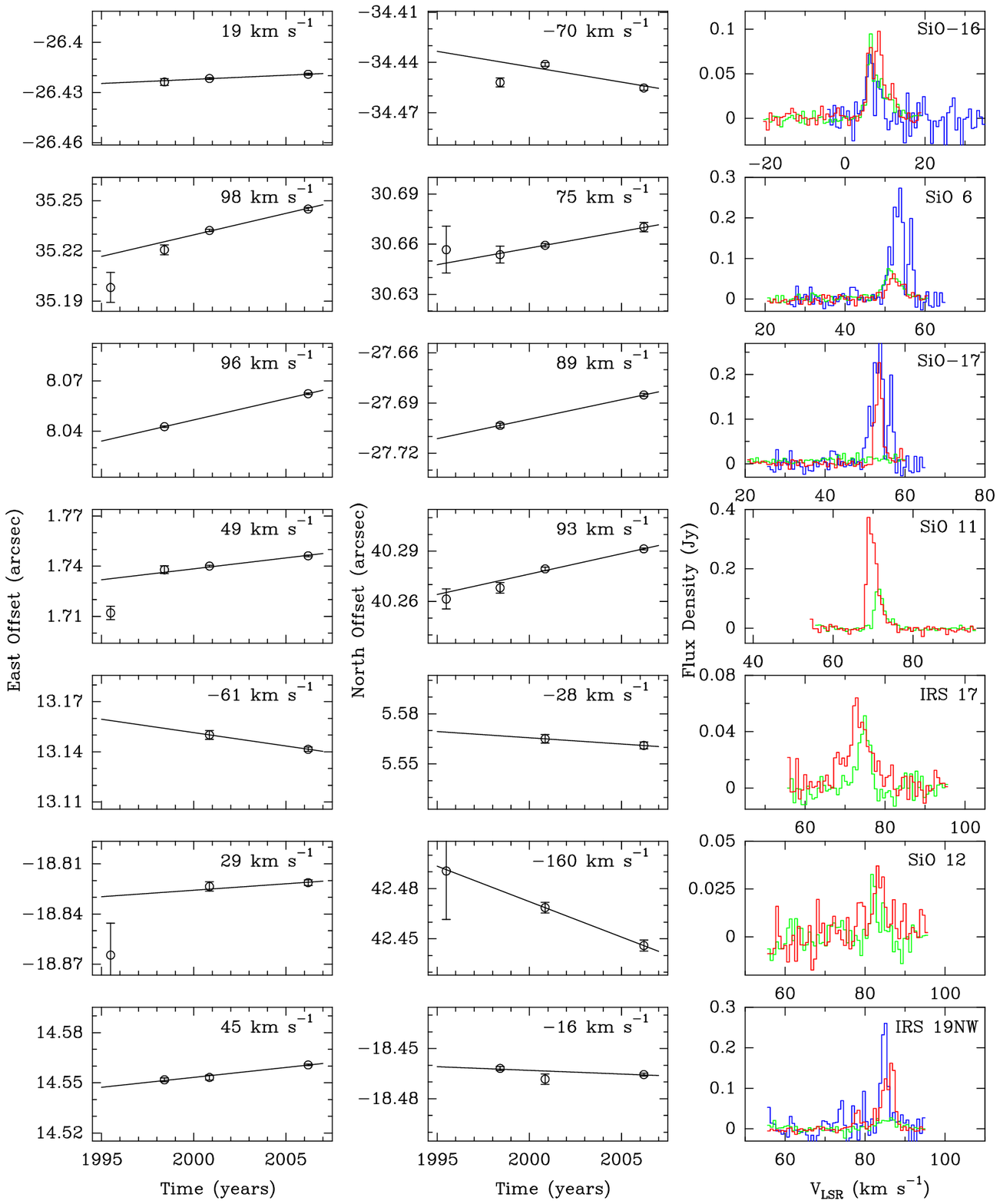} 
\caption{Eastward (left panels) and northward (middle panels) position
offsets from \sgra\ versus time for the seven SiO maser stars with
positive LSR velocities.  Solid lines are variance-weighted best-fit 
proper motions.  
The linear speed is indicated in each frame, assuming a distance of 8.0~kpc.  
Spectra from 1998 ({\it blue}), 2000 ({\it green}) and  
2006 ({\it red}) observations are also shown (right panels). 
Star names are indicated in the upper
right corner of the right panels.
            \label{fig:motions_b}
        }
\end{figure}

\begin{figure}
\epsscale{0.9}
\plotone {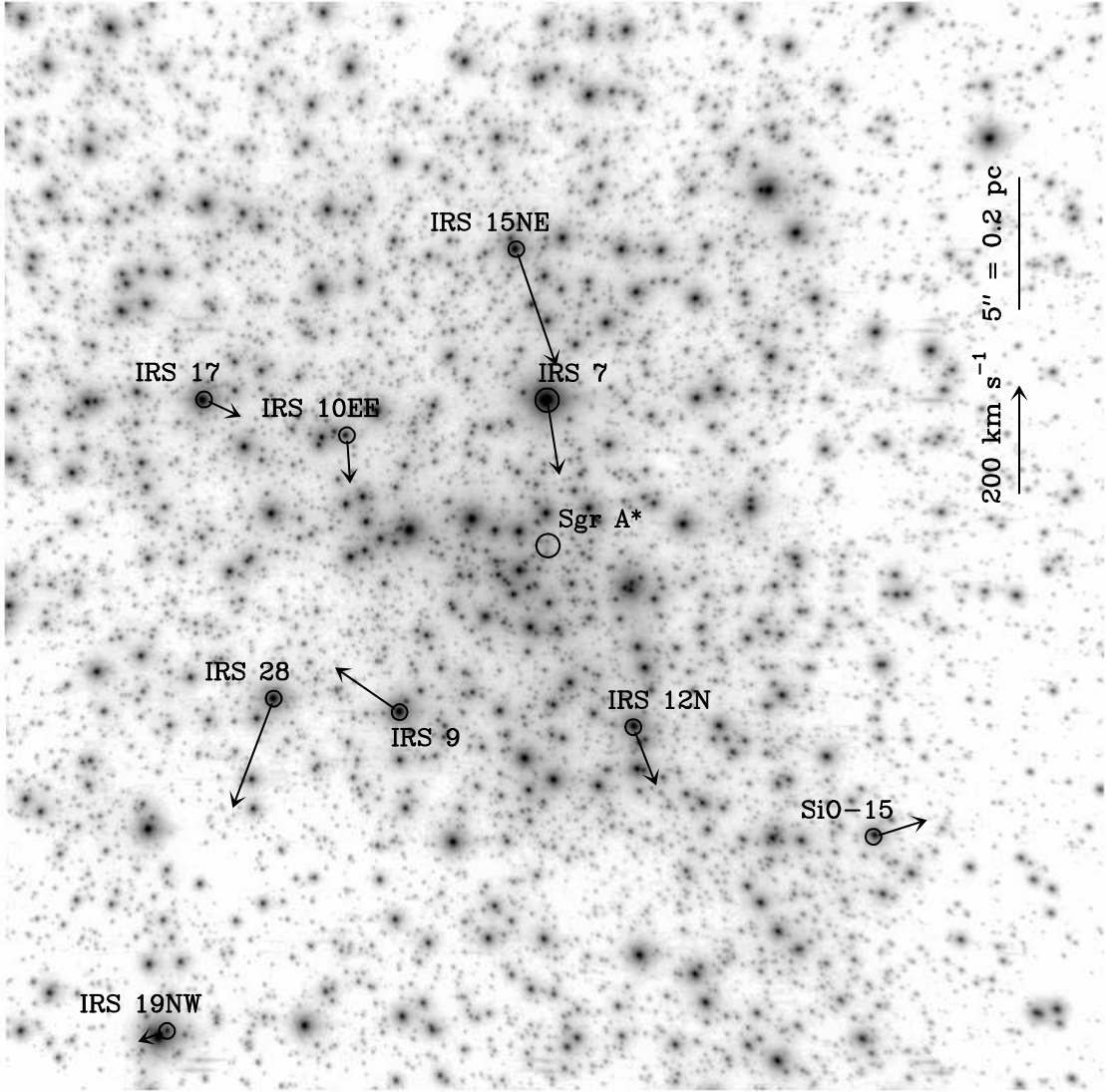} 
\caption{Infrared (K-band) image of the central $\pm20$'' of the Galactic Center
taken in 2005, with east to the left and north to the top.  
SiO maser stars within this region are circled and their 
proper motions relative to \sgra\ are indicated with arrows.  The vertical bar
and arrow at the right of the image indicate the linear and motion scales 
for $R_0=8.0$~kpc.  The location of \sgra\ is indicated by the
circle at the center of the image.
            \label{fig:ir_map}
        }
\end{figure}

We constructed spectra at the pixel of peak brightness for SiO masers
detected in the 1998, 2000, and 2006 VLA observations.  These spectra are
displayed in Figs.~\ref{fig:motions_a} \& \ref{fig:motions_b}.    
Most of these SiO spectra are as expected for Mira variables 
located at the distance of the the Galactic Center: they show flux 
densities $\lax1$~Jy covering a velocity range of $5~{\rm to}~10$~\kms\ 
and strong variability over timescales of years.  
Fig.~\ref{fig:ir_map} shows the Galactic Center
region, with the positions and proper motions of the nine SiO
maser stars that are projected within the inner 21'' displayed.

\subsection{SiO Maser Proper Motions}

We determined stellar proper motions by fitting a variance-weighted
straight line to
the positions as a function of time from all of the available data 
compiled in Table~\ref{table:positions}.  
These proper motion fits are given in Table~\ref{table:radio_motions} and 
displayed graphically in Figs.~\ref{fig:motions_a} \& \ref{fig:motions_b}.  
The reference epoch for the proper motion 
solution was chosen as the variance-weighted mean epoch for each star,
in order to obtain uncorrelated position and motion estimates.
Since the estimated uncertainties for individual east-west 
and north-south positions were neither identical, nor exactly linearly 
related, we chose a single, average reference epoch for each star 
(instead of a separate reference epoch for the east-west and 
north-south directions), which resulted in slight parameter correlations.

Assuming a distance of 8.0~kpc to the Galactic Center \citep{R93}, 
we convert proper motions to linear velocities and, with the radial
velocities, determine the full 3-dimensional speed, $V_{total}$, 
of each star with respect to \sgra.  
These speeds are given in Table~\ref{table:enclosed_masses}. 
While the speeds on the plane of the sky are directly referenced to
\sgra, the speeds along our line-of-sight are in the LSR reference
frame.  Thus, our values of $V_{total}$ assume that \sgra\ has a 
near-zero line-of-sight speed with respect to the LSR.  
As no spectral lines have been detected from \sgra, there is no direct 
observational evidence supporting this assumption.  
However, the lack of a detectable proper motion of \sgra\ 
suggests that it anchors the dynamical center of the Galaxy 
\citep{RB04} and should be nearly at rest in the LSR frame.

\section{Infrared Observations}

Near-IR observations were obtained on the 8.2-m UT4 (Yepun) of the ESO-VLT 
on Cerro Paranal, Chile, using the detector system NAOS/CONICA (hereafter NACO)
consisting of the adaptive optics system NAOS \citep{Rousset03} and the 
$1024 \times 1024$-pixel near-IR camera CONICA \citep{H03}.
We obtained 8 data sets in H and K bands with a scale of 
27 mas/pixel (large scale) covering 5 epochs (May 2002, May 2003, June 2004, 
May 2005, April 2006). 

Each image covers a field of view (FOV) of $28\times28$ arcsec. 
During each observation the camera pointing was shifted so that the total 
FOV of a complete data set was between $35\times35$ and $40\times40$~arcsec, 
centered on \sgra.  In all cases the spatial resolution was (nearly) diffraction 
limited, leading to a typical FWHM of $\sim60$ mas.  Typical limiting 
magnitudes were 18$^{th}$~mag at K band and 20$^{th}$ mag at H band.
All images were sky-subtracted, bad-pixel removed and flat-field corrected. 
In order to obtain the best signal-to-noise ratios and maximum 
FOV coverages in single maps, we combined all good-quality images belonging to 
the same data set to mosaics after correcting for instrumental geometric 
distortion.  Details of the distortion correction will be given in \citet{T06}. 

In order to establish an astrometric near-IR reference frame, 
we selected one high-quality 
mosaic (May 2005), extracted image positions for all detected stars and 
transformed them into astrometric coordinates using the positions of all 
9 maser stars in the FOV interpolated to the epoch of the image. 
To compare positions among images of different epochs, we chose an ensemble 
of $\approx600$ ``well-behaved'' stars (\ie stars that are well-separated from 
neighbors and are bright but not saturated) in the ``master mosaic''
and computed in each image all stellar positions relative to this ensemble. 
Due to varying FOVs, the number of stars usable for a given mosaic varied 
between about 400 to 600.

In each image, positions were extracted by fitting stars with 2-dimensional 
elliptical Gaussian brightness distributions.  Although over the entire 
FOV significant departures from isoplanicity occur, this effect elongates 
the stellar PSFs symmetrically and does not affect significantly the 
centroids of Gaussian-fitted positions.
Proper motions were computed by fitting the positions versus times with 
straight lines.  To assure statistically ``clean'' errors for the proper motions, 
outlier rejection and error rescaling were applied where possible. 
This led to typical position accuracies of $\approx1$~mas and typical proper 
motion accuracies of $\approx0.3$~\masy.
Unfortunately, this accuracy was not achieved for all maser stars,
as some are very bright stars and are affected by detector 
non-linearity/saturation in some images; 
also the star most distant from \sgra\ (IRS19NW) 
was observed only in the last two epochs. 
Thus, some of the maser stars have errors larger than typical.

\section{Radio \& Infrared Frame Alignments}

The proper motions of stars in the Galactic Center cluster, measured
from infrared images, are {\it relative} motions only.  One can
add an arbitrary constant vector to all of the stellar 
proper motions without violating observational constraints.
Until now, the ``zero points'' of the motions have been determined by 
assuming isotropy and removing the average motion of the entire
sample.   Since the radio proper motions are inherently in a
reference frame tied directly to \sgra, one can use any one of 
the SiO stars, or the mean motion of a group of them,
to place the infrared proper motions in \sgra's frame.

We have measured radio positions and proper motions, 
relative to \sgra, for the nine bright SiO maser stars that appear on
the NACO images (ie, within $\approx20$\arcs\ of \sgra).  
The position and proper motion accuracies
typically are $\sim1$~mas and $\sim0.3$~\masy, respectively.
This allows us to align the radio and infrared frames, both
in position and in proper motion.
Thus, not only can the location of \sgra\ can be accurately 
determined on infrared images, but also stellar proper motions
from infrared data can be referenced directly to \sgra, without 
assumptions of isotropy or homogeneity of the stellar motions.

The radio and infrared proper motions measured for the nine  
stars are listed in Table~\ref{table:motion_alignment}.  
The nine stars have weighted mean differences (and standard errors of 
the means) of  
$+0.66\pm0.21$~\masy\ toward the east, and
$-0.45\pm0.28$~\masy\ toward the north.
These results are qualitatively similar to those published in
Paper II.   Quantitatively, the differences between the IR motions
of Paper II and this paper for some stars (notably IRS 12N) are
greater than expected based on the quoted uncertainties.  Since
the IR motions in Paper II were based on 2 epochs only,
which did not allow for an internal check on the formal motion
uncertainties, we believe those uncertainties were somewhat
optimistic.

Currently only one star, IRS~7, has a significant discrepancy between the 
radio and infrared motions in the both coordinates.  This is the
only super-giant star in the sample and, owing to its extreme brightness,
the infrared measurements are compromised by detector saturation.
Additionally, the radio measurements are subject to 
significant uncertainty from the large SiO maser shell size. 
After removing IRS~7, the weighted mean differences between the
radio and IR motions change only slightly and become   
$+0.63\pm0.21$~\masy\ toward the east, and
$-0.32\pm0.18$~\masy\ toward the north.

When comparing how well the IR frame matches the radio frame,
we need to consider the statistical uncertainty of the average IR
motion, which has been removed.  For most epochs, the average
motion is based on $\approx400$ stars, each of which has a typical
motion of $\approx100$~\kms.  Thus, the mean IR motion should have
an uncertainty of roughly $100~\kms/\sqrt{400}~\approx 5$~\kms.  
Adopting the result with IRS~7 removed,
converting to linear speeds for a distance of 8.0~kpc to 
the Galactic Center \citep{R93}, and adding in quadrature a $\approx5$~\kms\ 
uncertainty for the mean IR motion removed from each coordinate, 
implies that the infrared stellar cluster moves
$+24\pm9$~\kms\ toward the east, and
$-12\pm9$~\kms\ toward the north, with respect to \sgra.
The northward component motion does not deviate significantly from zero;
the eastward component formally presents a $2.7\sigma$ significance. 
Combining these components in quadrature formally yields 
a speed difference of $27 \pm 9$~\kms.
However, at this time, we do not consider that we have firmly detected
motion of the stellar cluster, and we adopt
a $2\sigma$ upper limit of $45$~\kms\  
for the proper motion of the stellar cusp with respect to \sgra.

\section{Enclosed Mass versus Radius from \sgra}

Estimates of the enclosed mass versus projected radius from
\sgra, based on infrared stellar motions, rely on {\it relative} motions 
not tied directly to \sgra.
Since, the 3-dimensional motions of the SiO masers in this paper are
both very accurate and directly tied to \sgra, 
they provide valuable information on the enclosed mass within 
projected radii of 0.2 to 2 pc of \sgra.

In Paper II, we derived a lower limit to the enclosed mass at the
projected radius of each star, assuming the stellar motions reflect 
gravitational orbits dominated by a central point mass.  
Given the 3-dimensional speed, $V_{total}$, 
and projected distance from \sgra, $r_{proj}$, for each star, 
we obtained a strict lower limit to the mass enclosed, $M_{encl}$, 
within the true radius, $r$, of that star from \sgra.  
For a given enclosed mass, semi-major axis and 
eccentricity ($e$), the greatest orbital speed occurs at pericenter
for $e\approx1$.  
Since the {\it projected} pericenter distance cannot 
exceed the true distance, we obtained
$$M_{encl} \ge {V_{total}^2 r_{proj} \over 2G}~~.~~\eqno{(1)}$$
Note that this enclosed mass limit is a factor of two lower than
would be obtained for a circular orbit.
This lower limit approaches an equality only when three criteria 
are met: 1) $r_{proj} \approx r$, 2)
the star is near pericenter, and 3) it has an eccentricity near unity.
The {\it a priori} chance of any star satisfying all three of these criteria 
is small, especially since a star in a highly eccentric
orbit spends very little time near pericenter. 
Thus, Eq.~(1) provides a very conservative limit on enclosed mass.
 
We evaluate the lower limit to $M_{encl}$ using Eq.~1 
by adopting conservatively the smallest total velocity allowed by 
measurement uncertainties, \ie by subtracting $2\sigma$ from $V_{total}$
in Table~\ref{table:enclosed_masses} before calculating a mass limit.   
The mass limits, given in Table~\ref{table:enclosed_masses},
are mostly consistent with the enclosed mass versus projected distance from 
\sgra\ given by \citet{GEOE97} and \citet{G98}.
For many of the stars, the lower mass limits are well below the estimated
enclosed mass curves, as expected given the very conservative nature of the 
calculated limits.  

Our most significant lower mass limit is from IRS~9.
In Paper II, we arrived at a limit $>4.5\times10^6$~\msun, which
exceeded the then favored model of a $2.6\times10^6$~\msun\ 
black hole \citep{GEOE97,G98}, plus a $0.4\times10^6$~\msun\ contribution 
from the central stellar cluster \citep{G03}, by about 50\%.
With our improved proper motions, we now find a more stringent
limit of $>5.1\times10^6$~\msun\ at a projected radius of 0.33 pc from 
\sgra.  

\begin{figure}
\epsscale{0.95}
\plotone{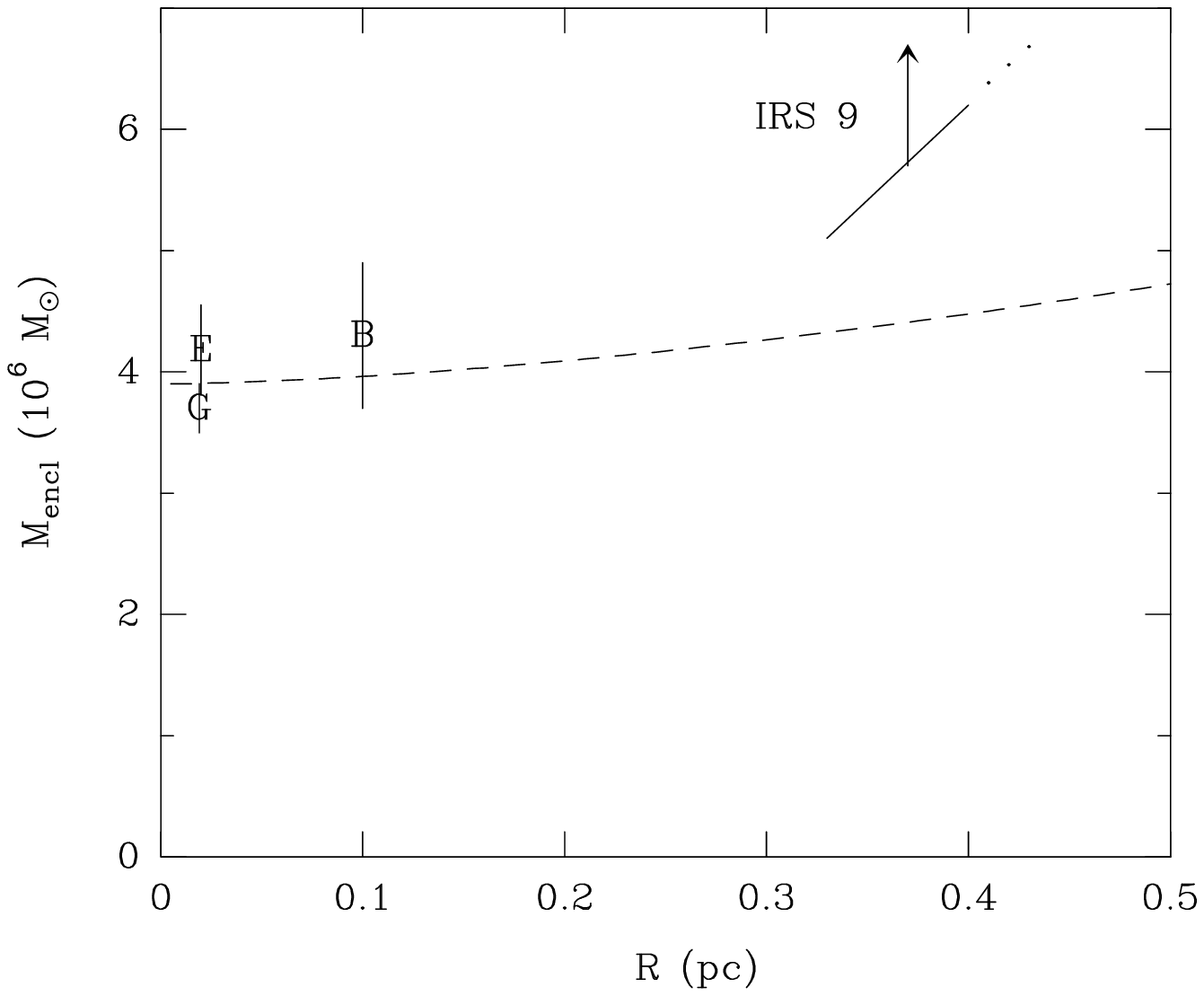}
\caption{Constraints on the enclosed mass as a function of radius ($R$) 
from \sgra.  The estimates labeled ``E'' and ``G'' are from
fitting stellar orbits by \citet{E05} and \citet{G05}, respectively,
and the estimate labeled ``B'' is from a statistical analysis of the 
``clockwise stellar disk'' by \citet{B06}.  
Our lower limit on enclosed mass from
the 3-dimensional motion of IRS~9, assuming that it is bound
in the region dominated gravitationally by Sgr~A* (see \S5.2), 
is indicated by a sloping line and arrow.
The minimum distance of IRS~9 from \sgra\ is its projected distance
of 0.33~pc.  The sloping line corresponds to the mass limit for
reasonable values of the unknown line-of-sight distance 
of IRS~9 from \sgra; this increases both the radius and the mass limit
linearly.  The dashed line indicates the combined contribution of
a point mass of $3.9\times10^6$~\msun\ and a central stellar cusp 
\citep{G03}.  
The uncertainty in the dashed line is dominated by the uncertainty
of $\pm0.2\times10^6$~\msun\ in the point-mass, with
a smaller contribution of perhaps $\pm0.1\times10^6$~\msun\ 
from the central stellar cusp at $R=0.33$~pc.
All values assume $R_0 = 8.0$~kpc; the discrepancy
between the dashed line and the IRS~9 limit cannot be removed for
values of $R_0 < 9$~kpc.  
            \label{fig:enclosed_mass}
        }
\end{figure}

Fig.~\ref{fig:enclosed_mass} displays our enclosed mass versus
radius constraint based on the 3-dimension motion of IRS~9, along with
other constraints in the recent literature.  
The current best estimate for the mass of the SMBH (\sgra)
is $(3.9\pm0.2)\times10^6$~\msun, for the distance to the Galactic Center, 
$R_0$, of $8.0$~kpc.   This mass estimate comes 
from an unweighted average of the results of \citet{E05} and \citet{G05},
based on stellar orbit determinations.
Adding in a $0.4\times10^6$~\msun\ contribution from the central stellar 
cluster, based on a density profile of 
$1.2\times10^6~(r/0.39~{\rm pc})^{-1.4}$~\msun~pc$^{-3}$ by \citet{G03},
yields $4.3\times10^6$~\msun, still leaving a discrepancy
of $0.8\times10^6$~\msun, for $R_0=8$~kpc.  
Formally, this is about a $3\sigma$ discrepancy, assuming an uncertainty
of $\pm0.2\times10^6$~\msun\ in the mass estimate of \sgra\ and 
an estimated $\pm30$\% uncertainty in the mass of the stellar cusp.  

Since we do not know the line-of-sight distance of IRS~9 from 
\sgra, one might be tempted to argue that $r \approx 2r_{proj}$
and the star is simply sensing an enclosed mass of $1.0\times10^6$~\msun\  
from the central stellar cluster at that radius.  However, the
mass limit derived from Eq.~(1) scales as $r$ and would be approximately
$10^7$~\msun\ for $r=2r_{proj}=0.66$~pc.  Thus, the mass discrepancy 
only {\it increases} for $r > r_{proj}$, as shown by the slanted line in
Fig.~\ref{fig:enclosed_mass} (but see \S5.2). 

How can the lower limit to the enclosed mass provided by IRS~9 be
explained?  We now discuss some possibilities.

\subsection{Dark Matter in the Central Stellar Cluster} 
One could explain the motion of IRS~9, were the central stellar
cluster to contain dark matter (in addition to \sgra) whose mass exceeds 
$0.8\times10^6$~\msun\ within $r=0.33$~pc.
\citet{M93} estimates that $\sim10^6$~\msun\ 
of ``dark'' stellar remnants (eg, white dwarfs, neutron stars, black holes)
could have accumulated in the inner few tenths of a parsec of the Galaxy.
\citet{M05} show that, with data available at the time, the orbital fit 
of star S2 allows for (but does not require) $0.2\times10^6$~\msun\ of 
dark matter distributed within 0.001~pc  of \sgra.
Should such a dark component exist and extend to 
greater radii, it might explain some of the IRS~9 mass discrepancy.
However, other estimates of the total mass in black holes in the
central few tenths of a pc do not exceed $\sim0.2\times10^6$~\msun\ 
\citep{MEG00,F06}.
Given these estimates and the evidence from 
other enclosed mass indicators that do not support $\sim10^6$~\msun\ of
dark matter within $\approx1$~pc of \sgra\ \citep{GPE00}, 
it seems unlikely that a dark component could explain more than a modest
fraction of the IRS~9 mass discrepancy.

\subsection{IRS~9 not bound to the central parsec}
A critical assumption for calculating the minimum enclosed mass
using IRS~9's space velocity (Eq.~1) is that it is in a bound orbit
dominated by a central point mass.   If IRS~9 is in a highly
eccentric orbit with a semimajor axis greater than a few parsecs,
this assumption can break down.  In such a case, the star's
space velocity could exceed the ``local'' escape velocity, based
on the mass enclosed at its current radius, but still be bound at 
a larger radius.  For example, a star could be bound by mass within 
$\approx10$~pc of \sgra, but observed plunging into the inner 
few tenths of a parsec at a speed that makes it appear unbound.
This could happen via gravitational scattering starting either at small
radii and increasing orbital energy or at large radii and removing
angular momentum.   \citet{vL92} suggest a similar explanation for
three OH/IR stars with large radial velocities; these stars are seen 
projected tens of parsecs from \sgra\ and would require 
semimajor axes of a few kpc.

Consider a Galactic Center star with a large semimajor axis
and little angular momentum ($e\approx1$), so that
it essentially ``falls'' toward the Center.  
We derive the infall velocity for a star that
starts ``falling'' at a radius, $r_{max}$ and reaches a radius of $r_0.$
Assume a central point mass, $M_{BH},$ plus an extended component
with density, $\rho(r).$
The kinetic energy gained by a star falling from $r$ to $r_0$ is
equal to the difference in gravitational potential energy at those radii.
Gravitational potential energy per unit mass, $U_m$, for a spherically 
symmetric mass distribution has the properties that for mass interior to $r$
$$U_m = -{G\over r}\int_0^r \rho(r)~4\pi r^2~dr~~~,$$
and for~mass~exterior~to $r$
$$U_m = -G \int_r^{r_{max}} {\rho(r)\over r}~4\pi r^2~dr~~~.$$
(For spherically symmetric systems, interior mass acts identically as a 
point mass equal to the enclosed mass located at the center of the 
distribution, while exterior mass results in zero gravitational force 
and a {\it constant} gravitational potential dependent on its radial position, 
but independent of the position of a ``test'' mass.)

Following \citet{G03}, $\rho(r) = \rho_0 (r/r_0)^\alpha~,$
where $\rho_0 = 1.2\times 10^6$~\msun~pc$^{-3}$, $r_0 = 0.39$~pc,
and $\alpha=-2.0$ for $r \ge r_0$~pc  and $\alpha\approx-1.4$ for $r < r_0$~pc.
This leads to an enclosed stellar mass within $r_0$ of $M_0/1.6,$
where $M_0 = 4\pi\rho_0 r_0^3 = 0.9\times10^6$~\msun, plus a contribution 
of $M_{BH}$ from \sgra.
Adding the contributions to the potential from different mass components
for a star at radius $r$ (for $r\ge r_0$) gives
$$U_m = -{G\over r}M_{BH} -{G\over r}\int_0^{r_0} \rho(r)~4\pi r^2~dr
        -{G\over r} \int_{r_0}^r \rho(r)~4\pi r^2~dr
        -G\int_r^{r_{max}} {\rho(r)\over r}~4\pi r^2~dr~~.~\eqno(2)$$
The first three terms on the right hand side of Eq.~(2) sum the
effects of the mass components interior to $r$, and the fourth term is the
contribution from the mass exterior to $r$.
Evaluating Eq.~(2) we find 
$$U_m = -{G\over r}M_{BH} -{G\over r}{M_0\over1.6}
        -{G\over r} M_0~({r\over r_0} - 1)
        -{G\over r_0} M_0~\biggl( \ln(r_{max}/r_0) - \ln (r/r_0) \biggr)
        ~~.~\eqno(3)$$
For a star ``falling'' from $r_{max}$ to $r_0$, the kinetic energy per unit mass
gained is equal to the difference in potential energy per unit mass.
From Eq.~(3), we find
$${1\over2} v^2 = U_m(r_{max}) - U_m(r_0)~~.~\eqno(4) $$
where
$$U_m(r_{max}) =  -{G\over r_{max}} \biggl(
                  M_{BH} + M_0 ({r_{max}\over r_0} - 0.38 )
                                  \biggr)~~~,$$
and
$$U_m(r_0) =  -{G\over r_0} \biggl(
                   M_{BH} + M_0 \bigl( 0.62 + \ln( r_{max}/r_0 ) \bigr)
                             \biggr)~~~,$$

Evaluating Eq.~(4) for $r_0 = 0.39$~pc, which is a reasonable value for
the 3-dimensional radius of IRS~9, gives $v > 370$~\kms\ for an
initial radius $r_{max} > 9$~pc.
Thus, if IRS~9 is in a highly eccentric orbit that takes it
out to a radius of $>9$~pc, it could achieve its very high 
observed 3-D velocity without violating the enclosed masses
estimated by other methods.

A priori it might seem very unlikely that even one of 15 stars with
detectable SiO masers would have such an orbit and be observed near
closest approach to \sgra\ (where it spends little time).  However, 
it is beyond the scope of this paper to evaluate the likelihood,
especially with the limited statistics available at this time. 
           
\subsection{The distance to the Galactic Center ($R_0$) exceeds $9$~kpc} 
Were the mass of \sgra\ $>4.7\times10^6$~\msun, no mass discrepancy would
exist.  The best current mass estimates are based on fitting  
orbits for many stars and should be robust.  However, the greatest 
uncertainty in the mass of \sgra\ comes its strong dependence on the 
adopted value of $R_0=8.0$~kpc for the distance to the Galactic Center.  
\citet{E05} derive central masses from orbit fitting of 
$4.06\times10^6$~\msun\ when adopting $R_0= 8.0$~kpc and
$3.61\times10^6$~\msun\ for a best fit $R_0= 7.62$~kpc.  
These values suggest an enclosed mass $M_{encl} \propto R_0^{2.4}$.
Our mass limit based on IRS~9's 3-D motion would also increase with $R_0$,
but more weakly.  Since the LSR velocity is the dominant component in 
the 3-dimensional motion for IRS~9 (and is not dependent on $R_0$), 
our minimum mass estimate (Eq.~1) scales approximately as 
$M_{encl} \propto R_0^{1.3}$, mostly through $r_{proj}.$ 
Allowing $R_0$ to increase to about 9~kpc removes the mass discrepancy.
However, such a large value for $R_0$ seems very unlikely \citep{R93,E05}.

\subsection{Non-zero $V_{\rm LSR}$ for \sgra} 
           
Were \sgra\ moving toward the Sun along the line-of-sight
with a speed $>30$~\kms,  this would lower $V_{total}$ 
and, hence, the $M_{encl}$ limit to $>4.4\times10^6.$
While it seems very unlikely that a super-massive object would
have such a large motion \citep{R03}, we now consider this possibility.
One method to approach this problem is to average the velocities
of large samples of stars very close to \sgra, assuming that this
average would apply to \sgra.

Our sample of SiO maser stars, which should be nearly complete
in the LSR velocity range $-131$ to $+95$~\kms, does not show
any obvious bias.  Infrared observations of CO band-head
velocities from late-type stars in the central parsecs
yield average LSR velocities that are not statistically different from zero.  
For example, the integrated CO-band head velocities (within a 20'' 
diameter aperture) of \citet{MSBH89} indicate positive (negative) velocities
at positive (negative) Galactic longitude, consistent with the direction
of Galactic rotation, and a value of $-10\pm25$~\kms\ at the position of \sgra.
(However, these authors find possibly significant stellar velocities of 
$-47\pm8$~\kms\ for four pointing offsets {\it perpendicular} to the Galactic
plane.)  Individual stellar velocities compiled by \citet{RR88} of 
54 stars projected within $\approx6$~pc of \sgra\ have a mean velocity 
of $-20\pm11$~\kms.  Alternatively, \citet{W85} and \citet{S98} 
measured velocities of OH masers for 33 and 229 OH/IR stars, respectively,
within about $\approx40$~pc of \sgra, which yield average velocities of 
$+7\pm11$~and $+4\pm5$ \kms.
Overall, it appears that radial velocities of stars near \sgra\  
suggest an average LSR velocity near zero, within $\approx20$~\kms.

\subsection{IRS~9 is (or was) in a binary}
Were IRS~9 in a tight, massive binary, perhaps a significant
portion of its space velocity might be contributed by internal 
orbital motion, possibly reducing its speed with respect to \sgra.  
However, we have observed IRS~9 for about 8 years and see no  
changes in its radial or proper motion velocity components.  
The spectra of IRS~9 shown in Fig.~\ref{fig:motions_a} are characteristic
of Mira variables, which show variable emission over a range of 
5 to 10~\kms\ about the stellar velocity.  We estimate that the
stellar radial velocity of IRS~9 has changed by less than 2~\kms\
over 8 years.   Also, the proper motion components are well-fit
by constant velocities, with $2\sigma$ upper limits to accelerations of
0.4 and 0.6 mas y$^{-2}$ (15 and 23~\kms~y$^{-1}$ at 8.0~kpc) 
for the eastward and northward components, respectively.  

The observed changes in radial velocity (or proper motion components)
would have different characteristics depending on the relative values 
of the time span of the observations, $\Delta t$, and the orbital period, 
$P$.  For $\Delta t \ge P/4$, we would have sampled large changes
in the orbital mean anomaly and hence would have seen quasi-random changes
of magnitude equal to the orbital speed. 
Our limit of $<2$~\kms\ change in the radial velocity over 8 y,
would place a limit of $\approx2$~\kms\ for the radial component
of any orbital velocity.  
Alternatively, if $\Delta t < P/4$, then we could be sampling only
a small portion of an orbit and detecting a velocity change
might be difficult.  However, setting $\Delta t=8$~y requires $P>32$~y.  
Since any bright companion for IRS~9 would have been observed,
we adopt companion mass $m$ of $\le 10$~\msun.  This would
even allow for most black hole companions.
For $m<10$~\msun\ and $P>32$~y, we find an upper limit
for an orbital speed of $<20$~\kms.
Thus, a binary orbital contribution to the observed space velocity
of IRS~9 could not exceed $\approx20$~\kms\ and likely would be 
considerably less.  Thus, it is highly unlikely that
the extreme velocity of IRS~9 could be explained as owing
to a binary orbit.

Could IRS~9 have been in a binary system and come unbound (or bound
at a much larger radius as discussed in \S5.2) after a close 
encounter with \sgra?  A small number of ``hyper-velocity'' stars are 
thought to have been ejected from the Galactic Center in this manner
\citep{H88,YT03,BG05}.  However, these are estimated to be very rare 
events ($<1$ in $10^5$ y) and we are statistically unlikely to be 
witnessing a newly created hyper-velocity star so close to \sgra.  
All hyper-velocity stars discovered to date are early-type 
main-sequence stars; they are found in the outer Galaxy and are 
moving at speeds of $\sim1000$~\kms, even after climbing out of the
gravitational potential of the inner Galaxy.  
Main sequence stars can survive the strong tidal forces 
experienced during close encounters with \sgra.  However, IRS~9 is an 
AGB star and, thus, is a very extended ($\approx300$ solar radii) 
and low surface-gravity object.  It is unclear if such a star could 
survive the ejection event, without losing its extended atmosphere.

\section {Conclusions}

We have now measured the radio positions and proper motions of 15 late
type stars with SiO maser emission in the Galactic Center stellar cluster.
All but two of these stars have been detected at three or more epochs
and have measurement accuracies of $\approx1$~mas in position and 
$\approx0.3$~\masy\ in proper motion.  Nine of these stars have multi-epoch
measurements of proper motions at infrared wavelengths.  A comparison
of the radio motions, which are relative to \sgra, with the
infrared motions indicates that the stellar cusp moves with \sgra\ 
to within 45~\kms.  

The three-dimensional speeds and projected distances of individual
stars from \sgra\ yield lower limits to the enclosed mass.
The enclosed mass limit for one star, IRS~9, exceeds current 
estimates of the combined mass of \sgra\ and the luminous stars
in the cusp within the central parsec.  
This result is puzzling, but might be explained, for example, by 
a combination of 
i) a population of dark stellar remnants in the central parsec, 
ii) IRS~9 being on a plunging ``orbit'' with a semimajor axis 
$\gg 1$~pc, and/or 
iii) a value of $R_0 > 8$~kpc.

\acknowledgments
We thank S. Gillessen for comments on the paper, and
A. Loeb and R. O'Leary for discussions on the extreme
motion of IRS~9.

\clearpage

\clearpage

\clearpage

\clearpage

\clearpage

\clearpage

\begin{deluxetable}{lrrrll}
\tabletypesize{\scriptsize}
\tablecaption   {SiO Maser Astrometry \label{table:positions}}
\tablehead{ \colhead{Star} &\colhead{$V_{\rm LSR}$} &\colhead{$\Delta\Theta_x$} &\colhead{$\Delta\Theta_y$} 
                           &\colhead{Epoch}     &\colhead{Telescope}   \\
            \colhead{}     &\colhead{(\kms)}    &\colhead{(arcsec)}         &\colhead{(arcsec)}      
                           &\colhead{(y)}       &\colhead{}
          }
\tablecolumns{6}
\startdata 
\\
IRS~9 \ .....     &$-342$  &$5.6515\pm0.0048$	 &$-6.3589\pm0.0060$    &1998.39   &VLA\\
		  &	   &$5.6501\pm0.0007$    &$-6.3509\pm0.0013$    &1998.41   &VLA\\
		  &	   &$5.6589\pm0.0011$    &$-6.3454\pm0.0017$    &2000.85   &VLA\\
		  &        &$5.6742\pm0.0004$    &$-6.3347\pm0.0009$	&2006.20   &VLA\\
\\
IRS~7 \ .....     &$-114$  &$0.0403\pm0.0080$    &$5.5829\pm0.0130$     &1995.49   &VLA\\   
                  &        &$0.0387\pm0.0023$    &$5.5676\pm0.0049$     &1998.39   &VLA\\
                  &        &$0.0378\pm0.0043$    &$5.5495\pm0.0014$     &1998.41   &VLA\\
                  &        &$0.0342\pm0.0016$    &$5.5414\pm0.0030$     &2000.85   &VLA\\
		  &	   &$0.0326\pm0.0007$	 &$5.5238\pm0.0013$     &2006.20   &VLA\\
\\
SiO-14\ .....    &$-112$  &$-7.6648\pm0.0012$   &$-28.4528\pm0.0020$	&1998.41   &VLA\\
		  &	   &$-7.6596\pm0.0005$   &$-28.4530\pm0.0008$   &2000.85   &VLA\\
		  &	   &$-7.6485\pm0.0005$	 &$-28.4586\pm0.0009$	&2006.20   &VLA\\
\\
IRS~12N\ ....     &$\p-63$ &$-3.2519\pm0.0005$    &$-6.8814\pm0.0005$ 	&1996.41   &VLBA\\
                  &        &$-3.2541\pm0.0005$    &$-6.8877\pm0.0006$ 	&1998.39   &VLA\\
                  &        &$-3.2543\pm0.0010$    &$-6.8860\pm0.0011$ 	&1998.41   &VLA\\
                  &        &$-3.2554\pm0.0009$    &$-6.8936\pm0.0012$ 	&2000.85   &VLA\\
	          &        &$-3.2626\pm0.0009$	  &$-6.9073\pm0.0019$	&2006.20   &VLA\\
\\
IRS~28\ .....     &$\p-55$ &$10.4702\pm0.0030$   &$-5.7884\pm0.0050$ 	&1998.41   &VLA\\
                  &        &$10.4693\pm0.0010$   &$-5.7956\pm0.0024$ 	&2000.85   &VLA\\
		  &	   &$10.4809\pm0.0007$	 &$-5.8254\pm0.0010$	&2006.20   &VLA\\
\\
SiO-15\ .....     &$\p-36$ &$-12.4253\pm0.0023$  &$-11.0794\pm0.0054$   &2000.85   &VLA\\
		  &	   &$-12.4385\pm0.0012$	 &$-11.0753\pm0.0015$	&2006.20   &VLA\\
\\
IRS~10EE\ ...     &$\p-27$ &$7.6821\pm0.0030$    &$4.2125\pm0.0050$ 	&1995.49   &VLA\\
                  &        &$7.6837\pm0.0005$    &$4.2194\pm0.0005$ 	&1996.41   &VLBA\\
                  &        &$7.6841\pm0.0005$    &$4.2146\pm0.0009$ 	&1998.39   &VLA\\
                  &        &$7.6837\pm0.0005$    &$4.2157\pm0.0005$ 	&1998.41   &VLA\\
                  &        &$7.6845\pm0.0005$    &$4.2099\pm0.0005$ 	&2000.85   &VLA\\
		  &	   &$7.6840\pm0.0005$	 &$4.1990\pm0.0005$	&2006.20   &VLA\\
\\
IRS~15NE\ ...     &$\p-12$ &$1.2256\pm0.0140$    &$11.3108\pm0.0210$    &1995.49   &VLA\\
                  &        &$1.2302\pm0.0005$    &$11.3315\pm0.0005$    &1996.41   &VLBA\\
                  &        &$1.2249\pm0.0017$    &$11.3193\pm0.0019$    &1998.39   &VLA\\
                  &        &$1.2270\pm0.0005$    &$11.3201\pm0.0006$    &1998.41   &VLA\\
                  &        &$1.2228\pm0.0011$    &$11.3024\pm0.0025$    &2000.85   &VLA\\
		  &	   &$1.2112\pm0.0005$	 &$11.2761\pm0.0010$	&2006.20   &VLA\\
\\
SiO-16\ .....     &$\p\p+8$&$-26.4237\pm0.0020$	 &$-34.4520\pm0.0027$	&1998.41   &VLA\\
		  &	   &$-26.4216\pm0.0006$  &$-34.4412\pm0.0011$   &2000.85   &VLA\\
		  &	   &$-26.4191\pm0.0006$	 &$-34.4553\pm0.0013$	&2006.20   &VLA\\
\\
SiO--6\ .....     &$\p+52$ &$35.1982\pm0.0090$   &$30.6567\pm0.0140$    &1995.49   &VLA\\
                  &        &$35.2207\pm0.0029$   &$30.6537\pm0.0050$    &1998.39   &VLA\\
                  &        &$35.2323\pm0.0006$   &$30.6593\pm0.0010$    &2000.85   &VLA\\
	 	  &	   &$35.2451\pm0.0010$	 &$30.6702\pm0.0028$	&2006.20   &VLA\\
\\
SiO-17\ .....     &$\p+53$ &$8.0427\pm0.0005$    &$-27.7034\pm0.0014$   &1998.41   &VLA\\
                  &        &$8.0624\pm0.0005$    &$-27.6852\pm0.0009$   &2006.20   &VLA\\
\\
SiO--11\ ....     &$\p+70$ &$1.7121\pm0.0040$    &$40.2614\pm0.0060$    &1995.49   &VLA\\
                  &        &$1.7379\pm0.0023$    &$40.2681\pm0.0032$    &1998.39   &VLA\\
                  &        &$1.7401\pm0.0005$    &$40.2794\pm0.0011$    &2000.85   &VLA\\
		  &	   &$1.7462\pm0.0005$	 &$40.2914\pm0.0006$	&2006.20   &VLA\\
\\
IRS~17\ .....     &$\p+73$ &$13.1501\pm0.0026$   &$5.5651\pm0.0025$     &2000.85   &VLA\\
		  &	   &$13.1415\pm0.0013$	 &$5.5611\pm0.0021$	&2006.20   &VLA\\
\\
SiO--12\ ....     &$\p+82$ &$-18.8645\pm0.0190$   &$42.4905\pm0.0290$   &1995.49   &VLA\\
                  &        &$-18.8235\pm0.0028$   &$42.4686\pm0.0032$   &2000.85   &VLA\\
	 	  &	   &$-18.8212\pm0.0017$	  &$42.4459\pm0.0033$	&2006.20   &VLA\\
\\
IRS~19NW\ .....     &$\p+84$ &$14.5518\pm0.0011$	 &$-18.4619\pm0.0012$   &1998.41   &VLA\\
		  &	   &$14.5532\pm0.0015$   &$-18.4683\pm0.0031$   &2000.85   &VLA\\
		  &	   &$14.5607\pm0.0005$	 &$-18.4656\pm0.0009$	&2006.20   &VLA\\
\\
\tablecomments{VLBA positions are reported at a single reference epoch.
VLA data have been corrected for differential aberration.
$\Delta\Theta_x$ and $\Delta\Theta_y$ are angular offsets,
and $\mu_x$ and $\mu_y$ are proper motions, relative to  
\sgra\ toward the east and north, respectively, in the J2000 
coordinate system.
}
\enddata
\end{deluxetable}

\clearpage

\begin{deluxetable}{lrrrrrcc}
\tabletypesize{\scriptsize}
\tablecaption   {SiO Maser Proper Motions \label{table:radio_motions}}
\tablehead{ \colhead{Star} &\colhead{$V_{\rm LSR}$} &\colhead{$\Delta\Theta_x$} &\colhead{$\Delta\Theta_y$}
                           &\colhead{$\mu_x$}   &\colhead{$\mu_y$}
                           &\colhead{Epoch}     &\colhead{Number} \\
            \colhead{}     &\colhead{(\kms)}    &\colhead{(arcsec)}              &\colhead{(arcsec)}      
                           &\colhead{(\masy)} &\colhead{(\masy)}&\colhead{(y)} &\colhead{Epochs}   
          }
\tablecolumns{8}
\startdata 
\\
IRS~9 \ .....     &$-342$  &$+5.6655\pm0.0003$    &$-6.3407\pm0.0007$   &$+3.06\pm0.10$ &$+2.11\pm0.19$ &2003.34   &3 \\
IRS~7 \ .....     &$-114$  &$+0.0336\pm0.0050$    &$+5.5401\pm0.0050$   &$-0.58\pm0.50$ &$-3.52\pm0.54$ &2004.37   &4 \\
SiO-14\ ....     &$-112$  &$-7.6554\pm0.0003$    &$-28.4553\pm0.0006$  &$+2.08\pm0.12$ &$-0.94\pm0.20$ &2002.89   &3 \\
IRS~12N\ ....     &$\p-63$ &$-3.2537\pm0.0003$    &$-6.8864\pm0.0003$ 	&$-1.06\pm0.10$ &$-2.70\pm0.17$ &1998.17   &4 \\
IRS~28\ .....     &$\p-55$ &$+10.4784\pm0.0011$   &$-5.8190\pm0.0010$ 	&$+2.00\pm0.38$ &$-5.29\pm0.42$ &2005.00   &3 \\
SiO-15\ .....     &$\p-36$ &$-12.4372\pm0.0011$   &$-11.0757\pm0.0015$  &$-2.47\pm0.98$ &$+0.77\pm2.10$ &2005.68   &2 \\
IRS~10EE\ ...     &$\p-27$ &$+7.6840\pm0.0003$    &$+4.2114\pm0.0003$ 	&$+0.04\pm0.07$ &$-2.09\pm0.07$ &2000.24   &5 \\
IRS~15NE\ ...     &$\p-12$ &$+1.2257\pm0.0003$    &$+11.3171\pm0.0004$  &$-1.96\pm0.07$ &$-5.68\pm0.12$ &1998.92   &5 \\
SiO-16\ .....     &$\p\p+8$&$-26.4207\pm0.0004$   &$-34.4478\pm0.0043$  &$+0.49\pm0.15$ &$-1.84\pm1.52$ &2002.82   &3 \\
SiO--6\ .....     &$\p+52$ &$+35.2333\pm0.0011$   &$+30.6605\pm0.0009$  &$+2.58\pm0.43$ &$+1.99\pm0.52$ &2001.43   &4 \\
SiO-17\ .....     &$\p+53$ &$+8.0560\pm0.0004$    &$-27.6911\pm0.0008$  &$+2.53\pm0.18$ &$+2.34\pm0.42$ &2003.68   &2 \\
SiO--11\ ....     &$\p+70$ &$+1.7441\pm0.0014$    &$+40.2871\pm0.0007$  &$+1.30\pm0.46$ &$+2.45\pm0.25$ &2004.38   &4 \\
IRS~17\ .....     &$\p+73$ &$+13.1442\pm0.0012$   &$+5.5624\pm0.0016$   &$-1.61\pm1.08$ &$-0.75\pm1.22$ &2004.49   &2 \\
SiO--12\ ....     &$\p+82$ &$-18.8227\pm0.0030$   &$+42.4559\pm0.0023$  &$+0.77\pm1.14$ &$-4.24\pm0.84$ &2003.83   &3 \\
IRS~19NW \ ..     &$\p+84$ &$+14.5578\pm0.0005$   &$-18.4647\pm0.0012$  &$+1.19\pm0.14$ &$-0.43\pm0.31$ &2003.79   &3 \\
\\
\tablecomments{  
For sources with VLBA detections, only a single position was used when fitting
for proper motions.  VLA data have been corrected for differential aberration.
$\Delta\Theta_x$ and $\Delta\Theta_y$ are angular offsets at the listed epoch,
and $\mu_x$ and $\mu_y$ are proper motions, relative to  
\sgra\ toward the east and north, respectively, in the J2000 
coordinate system.
IRS~7 was shifted by +0.010'' northward to ``center the star'' between two maser 
positions; its position and proper motion uncertainties were increased to 0.005'' 
and 0.5~\masy.
Formal proper motion uncertainties were doubled for the stars with only 2-epoch
motions.
}
\enddata
\end{deluxetable}

\clearpage

\begin{deluxetable}{lrrrrrrrcc}
\tabletypesize{\scriptsize}
\tablecaption   {3-Dimensional Stellar Motions \& Enclosed Mass Limits
                 \label{table:enclosed_masses} }
\tablehead{ \colhead{Star} &&\colhead{$V_{\rm LSR}$} 
           &\colhead{$V_x$}     &\colhead{$V_y$}     
           &&\colhead{$V_{total}$}     
	   &&\colhead{R$_{proj}$}          &\colhead{M$_{encl}$}     \\
            \colhead{}     &&\colhead{(\kms)}    
           &\colhead{(\kms)}    &\colhead{(\kms)}   
           &&\colhead{(\kms)}         
           &&\colhead{(pc)}            &\colhead{($10^6$~\msun)}
           }
\tablecolumns{10}
\startdata 
\\
IRS~9   \ ...  &&$-342\pm  3$  &$ 116\pm\p4$  &$  80\pm\p7$  &&$ 370\pm\p3$  &&$ 0.33$ &$>5.1$ \\
IRS~7   \ ...  &&$-114\pm  3$  &$ -22\pm 19$  &$-133\pm 20$  &&$ 177\pm 16$  &&$ 0.21$ &$>0.5$ \\
SiO-14 \ ...  &&$-112\pm  3$  &$  79\pm\p5$  &$ -36\pm\p8$  &&$ 142\pm\p4$  &&$ 1.14$ &$>2.4$ \\
IRS~12N \ ...  &&$ -63\pm  3$  &$ -40\pm\p4$  &$-102\pm\p6$  &&$ 127\pm\p6$  &&$ 0.30$ &$>0.5$ \\
IRS~28  \ ...  &&$ -55\pm  3$  &$  76\pm 14$  &$-201\pm 16$  &&$ 221\pm 15$  &&$ 0.46$ &$>2.0$ \\
SiO-15  \ ...  &&$ -36\pm  3$  &$ -94\pm 37$  &$  29\pm 80$  &&$ 105\pm 40$  &&$ 0.65$ &$>0.0$ \\
IRS~10EE\ ...  &&$ -27\pm  3$  &$  -5\pm\p3$  &$ -82\pm\p3$  &&$  87\pm\p3$  &&$ 0.34$ &$>0.3$ \\
IRS~15NE\ ...  &&$ -12\pm  3$  &$ -74\pm\p3$  &$-215\pm\p5$  &&$ 228\pm\p4$  &&$ 0.44$ &$>2.5$ \\
SiO-16  \ ...  &&$   8\pm  3$  &$  19\pm\p6$  &$ -70\pm 58$  &&$  73\pm 55$  &&$ 1.68$ &$>0.0$ \\
SiO--6  \ ...  &&$  52\pm  3$  &$  98\pm 16$  &$  75\pm 20$  &&$ 134\pm 16$  &&$ 1.81$ &$>2.2$ \\
SiO-17  \ ...  &&$  53\pm  3$  &$  96\pm\p3$  &$  89\pm\p8$  &&$ 141\pm\p6$  &&$ 1.12$ &$>2.2$ \\
SiO--11 \ ...  &&$  70\pm  3$  &$  49\pm 17$  &$  93\pm\p9$  &&$ 126\pm 10$  &&$ 1.56$ &$>2.1$ \\
IRS~17  \ ...  &&$  73\pm  3$  &$ -61\pm 41$  &$ -28\pm 46$  &&$  99\pm 29$  &&$ 0.55$ &$>0.1$ \\
SiO--12 \ ...  &&$  82\pm  3$  &$  29\pm 43$  &$-160\pm 32$  &&$ 183\pm 29$  &&$ 1.80$ &$>3.3$ \\
IRS~19NW\ ...  &&$  84\pm  3$  &$  45\pm\p5$  &$ -16\pm 12$  &&$  97\pm\p4$  &&$ 0.91$ &$>0.8$ \\
\tablecomments{
$V_x$ and $V_y$ are proper motions speeds toward the East and North, 
respectively.
$V_{total} = \sqrt{V_{\rm LSR}^2 + V_x^2 + V_y^2}$ is the total speed of
the stars relative to \sgra.  
Proper motion speeds, projected distances, total speeds and enclosed mass
limits assume a distance to the Galactic Center of 8.0~kpc.
}
\enddata
\end{deluxetable}

\clearpage

\begin{deluxetable}{lrrrrrrrr}
\tabletypesize{\scriptsize}
\tablecaption   {Radio--Infrared Proper Motion Alignment
                 \label{table:motion_alignment} }
\tablehead{ \colhead{Star}  
           &\colhead{$\mu_x^{Radio}$} &\colhead{$\mu_y^{Radio}$} 
           &&\colhead{$\mu_x^{IR}$}    &\colhead{$\mu_y^{IR}$} 
           &&\colhead{$\mu_x^{Dif}$}    &\colhead{$\mu_y^{Dif}$} 
     \\
            \colhead{}      
           &\colhead{(\masy)}         &\colhead{(\masy)}      
           &&\colhead{(\masy)}         &\colhead{(\masy)}      
           &&\colhead{(\masy)}         &\colhead{(\masy)}      
          }
\tablecolumns{9}
\startdata 
\\
IRS9    \ ...  &$ 3.06\pm0.10$  &$ 2.11\pm0.19$  &&$ 4.44\pm0.55$  &$ 1.10\pm0.58$  &&$ 1.38\pm0.56$  &$-1.01\pm0.61$ \\
IRS7    \ ...  &$-0.58\pm0.50$  &$-3.52\pm0.54$  &&$ 1.86\pm1.18$  &$-7.33\pm0.72$  &&$ 2.44\pm1.28$  &$-3.81\pm0.90$ \\
IRS12N  \ ...  &$-1.06\pm0.10$  &$-2.70\pm0.17$  &&$-0.77\pm0.66$  &$-3.39\pm0.39$  &&$ 0.29\pm0.67$  &$-0.69\pm0.43$ \\
IRS28   \ ...  &$ 2.00\pm0.38$  &$-5.29\pm0.42$  &&$ 2.27\pm0.35$  &$-5.85\pm0.31$  &&$ 0.27\pm0.52$  &$-0.56\pm0.52$ \\
SiO-15  \ ...  &$-2.47\pm0.98$  &$ 0.77\pm2.10$  &&$-2.18\pm0.51$  &$-0.56\pm0.12$  &&$ 0.29\pm1.10$  &$-1.33\pm2.10$ \\
IRS10EE \ ...  &$ 0.04\pm0.08$  &$-2.09\pm0.07$  &&$ 0.73\pm0.23$  &$-1.92\pm0.27$  &&$ 0.69\pm0.24$  &$ 0.17\pm0.28$ \\
IRS15NE \ ...  &$-1.96\pm0.07$  &$-5.68\pm0.12$  &&$-2.40\pm0.48$  &$-6.29\pm0.35$  &&$-0.44\pm0.49$  &$-0.61\pm0.37$ \\
IRS17   \ ...  &$-1.61\pm1.08$  &$-0.75\pm1.22$  &&$ 0.20\pm0.59$  &$-1.67\pm0.62$  &&$ 1.81\pm1.23$  &$-0.92\pm1.37$ \\
IRS19NW \ ...  &$ 1.19\pm0.14$  &$-0.43\pm0.31$  &&$-0.60\pm3.08$  &$-0.54\pm3.47$  &&$-1.79\pm3.08$  &$-0.11\pm3.48$ \\
\\
\tablecomments{$\mu_x$ and $\mu_y$ are proper motions relative to \sgra\
toward the east and north, respectively.  Differenced motions 
(infrared minus radio) are indicated with the superscript ``Dif''.
Radio motions are in a reference frame tied to \sgra; infrared motions
are relative motions, with an average of $\approx400$ star motions removed.
}
\enddata
\end{deluxetable}

\end{document}